\documentclass[preprint,showpacs,preprintnumbers,amsmath,amssymb]{revtex4}
\usepackage{graphicx}
\usepackage{bm}
\usepackage{dcolumn}
\begin{document}
\begin{abstract}
We investigate the spin dynamics and the conservation of helicity
 in the first order $S-$matrix of a Dirac particle in any  static
magnetic field. We express the dynamical quantities using a
coordinate system defined by the three mutually orthogonal
vectors; the total momentum $\mathbf{k}=\mathbf{p_f}+\mathbf{p_i}$
, the momentum transfer $\mathbf{q}=\mathbf{p_f-p_i}$, and
$\mathbf{l}=\mathbf{k\times q}$. We show that this leads to an
alternative symmetric description of the conservation of helicity
in a static magnetic field at first order. In particular, we show
that helicity conservation in the transition can be viewed as the
invariance of the component of the spin along $\mathbf{k}$, and
the flipping of its component along $\mathbf{q}$, just as what
happens to the momentum vector of a ball bouncing off a wall. We
also derive a "plug and play" formula for  the transition matrix
element where the only reference to the specific field
configuration , and the incident and outgoing momenta is through
the kinematical factors multiplying a general matrix element that
is independent of the specific vector potential present.
\end{abstract}
\title{$SU(2)$ Symmetry and Conservation of Helicity for a Dirac Particle in
a Static Magnetic Field at  First Order}
\author{M.S.Shikakhwa\footnote{Present address: Physics Program, Middle East Technical University NCC, via Mersin 10,
Turkey}}
\email{moody@ju.edu.jo}%
\address{Department of Physics, University of Jordan\\
11942--Amman, Jordan}%
\author{A.Albaid}
\address{Department of Physics, Oklahoma Sate University
145 Physics Building,Stillwater OK 74078-3072 USA,}
\email{abdelhamid.albaid@okstate.edu}

\maketitle


\maketitle
\section{Introduction}
The use of the helicity, i.e.the projection of the spin along the
direction of the momentum, to describe the polarization of Dirac
particles in collision problems became common as a result of the
pioneering work  by Jacob and Wick \cite{Jacob}. Obviously, the
reason is that the energy eigenstates of the Hamiltonian are also
helicity eigenstates. In particular the plane wave solutions of
the free Dirac equation which are used to represent the incident
and outgoing particles in the first order $S-$matrix are
simultaneous eigenstates of the helicity operator
$\mathbf{\Sigma}.\mathbf{P}$ of the particle . The analysis of
collisions with the use of these basis is greatly simplified. \\
Among the interactions that conserve helicity, probably, the
interaction with a  static magnetic field is the most popular. As
 is well-known, the helicity of a Dirac particle in an electromagnetic potential is
conserved given that there is no electric field acting on the
particle \cite{sakurai}. Indeed, the Heisenberg equation of motion
for the helicity operator $\mathbf{\Sigma}.\mathbf{\Pi}$ where
$\mathbf{\Pi}=(\mathbf{p}-e\mathbf{A})$ is the mechanical momentum
of the particle reads ($\hbar=c=1$):
\begin{equation}\label{1}
[ \mathbf{\Sigma}.\mathbf{\Pi},H]=e\mathbf{\Sigma}\cdot\mathbf{E}
\end{equation}
Here, $H$ is the Hamiltonian of a Dirac particle in an
electromagnetic field. Thus, the  helicity of a particle in a
static magnetic field is conserved. In physical terms,
conservation of helicity is described as the invariance of the
component  of the spin of the particle along its momentum. In the
perturbative expansion of a helicity-conserving theory, helicity
is conserved at each order of the perturbation series . For
example, in the first order S-matrix element of the elastic
scattering of a particle in some helicity-conserving vector
potential, the conservation of helicity manifests itself through
the fact that if the incident state is in an eigen state of the
helicity operator $\mathbf{\Sigma}.\mathbf{\hat{p}}_i$
($\mathbf{\hat{p}}_i\equiv\frac{\mathbf{p}_i}{|\mathbf{p}_i|}$),
then the matrix element for the transition to a final state with
the opposite helicity is zero \cite{sakurai} ( $\mathbf{p}_i$ and
$\mathbf{p}_f$ are,respectively, the incident and outgoing
momenta). This work focuses on the conservation of helicity for
the scattering of a Dirac particle in a static magnetic field
\textit{at this order}. It is shown that , by formulating the
whole spin dynamics in terms of the three operators
$\Sigma_k=\mathbf{\Sigma}.\mathbf{\hat{k}};\quad\Sigma_q=\mathbf{\Sigma}.\mathbf{\hat{q}}
$ and $\;\Sigma_l=\mathbf{\Sigma}.\mathbf{\hat{l}}$, with the
three mutually orthogonal vectors; the total momentum
$\mathbf{k}=\mathbf{p_f}+\mathbf{p_i}$ , the momentum transfer
$\mathbf{q}=\mathbf{p_f-p_i}$, and $\mathbf{l}=\mathbf{k\times
q}$, one gets a more symmetric and intuitive picture of the
dynamics that lead to the conservation of the helicity in the
transition. It is also demonstrated that one can, within this
framework, express  the  helicity sector of the the matrix element
in a form that is independent of the specific  form of the vector
potential.
\section{ The Spin Interaction}
Consider a Dirac particle in a given magnetic field whose vector
potential is the static vector function $\mathbf{A\mathbf(x)}$ and
such that there is no scalar potential. The first order S-matrix
element for the elastic scattering of a particle in this potential
is :
\begin{equation}
\label{5}S_{fi}^{(1)}= i\int {d^4 x\,\bar \psi _f \left( x
\right)\left( e\mathbf{\gamma}\cdot\mathbf{A} \right)\psi _i
\left( x \right)}.
\end{equation}
Carrying out the time integral,we get this as
\begin{equation}
\label{5b}S_{fi}^{(1)}=- 2\pi e|N|^{2} \delta (E_f-E_i)
u^{\dagger} _f \left( p_f,s_f \right)\left(\int {d^3
x\,e^{i\left(\mathbf{p}_f-\mathbf{p}_i\right)\cdot\mathbf{x}}\left(
\mathbf{\alpha}\cdot \mathbf{A} \right)}\right)u _i \left(p_i,s_i
\right).
\end{equation}
which can be casted in the form
\begin{equation}
\label{5c}S_{fi}^{(1)}=- 2\pi e|N|^{2} \delta (E_f-E_i)
u^{\dagger} _f \left( p_f,s_f \right)\left( \mathbf{\alpha}\cdot
\mathbf{A}(\mathbf{q}) \right)u _i \left(p_i,s_i \right).
\end{equation}
where $\mathbf{A}(\mathbf{q})$ is the Fourier transform of the
vector potential with respect to the momentum transfer vector
$\mathbf{q}=\mathbf{p}_f-\mathbf{p}_i$ and $N$ is a normalization
constant . Recalling that $\alpha_i=\gamma_5\Sigma_i$, where
$\Sigma_i=\frac{i}{2}[\gamma_i,\gamma_j]\qquad ,\quad (i,j=1..3)$,
and $i\gamma_5=\gamma_1\gamma_2\gamma_3\gamma_4$, with $\gamma$'s
being the Dirac matrices $\{\gamma_\mu,\gamma_\nu \}=2g_{\mu\nu}$
,we write the matrix element as:
\begin{equation}
 \label{5}S_{fi}^{(1)}=- 2\pi e|N|^{2}|\mathbf{A}(\mathbf{q})| \delta
(E_f-E_i) u^{\dagger} _f \left( p_f,s_f \right)\left(
\gamma_5\mathbf{\Sigma}.\mathbf{\hat{a}} \right)u _i \left(p_i,s_i
\right).
\end{equation}
where we have introduced the unit vector
$\mathbf{\hat{a}}=\frac{\mathbf{A}(\mathbf{q})}{|\mathbf{A}(\mathbf{q})|}$.
The operator $\gamma_5\mathbf{\Sigma}.\mathbf{\hat{a}}$ is what we
denote with the spin
 interaction operator (SI) as it is the operator that induces transition in the spin space of
 the particle. The Helicity conservation  is
 reflected in the first order transition  as the vanishing of the helicity flip
 scattering matrix element ;
 \begin{equation}
 \label{5b}S_{fi}^{(1)}=- 2\pi e|N|^{2}|\mathbf{A}(\mathbf{q})| \delta
(E_f-E_i) u^{\dagger} _f \left( p_f,\mp \right)\left(
\gamma_5\mathbf{\Sigma}.\mathbf{\hat{a}} \right)u _i \left(p_i,\pm
\right)=0.
\end{equation}
where $u _i \left(p_i,\pm\right)$ are the eigenstates of
$\mathbf{\Sigma}.\mathbf{\hat{p}}_i$ with eigenvalues $\pm 1$. We
will focus now on  the non-vanishing spin-space matrix element
$\mathcal{M}$, and express it using the Dirac notation:
 \begin{equation}
 \label{5c}\mathcal{M}= u^{\dagger} _f \left( p_f,\pm \right)\left(
\gamma_5\mathbf{\Sigma}.\mathbf{\hat{a}} \right)u _i \left(p_i,\pm
\right)\\
=<\mathbf{\hat{p_f}};\pm|\gamma_5\mathbf{\Sigma}.\mathbf{\hat{a}}|\mathbf{\hat{p_i}};\pm>
\end{equation}
We now note that the two unit vectors;
$\mathbf{\hat{k}}=\frac{\mathbf{p_f}+\mathbf{p_i}}
{|\mathbf{p_f}+\mathbf{p_i}|}$ along the total momentum and
 $\mathbf{\hat{q}}=\frac{\mathbf{p_f}-\mathbf{p_i}}
{|\mathbf{p_f}-\mathbf{p_i}|}$ along the momentum transfer are
orthonormal; see Figure 1. This is, of course, true for the
scattering in any potential field.
\begin{figure}[htbp]
\includegraphics[width=8cm]{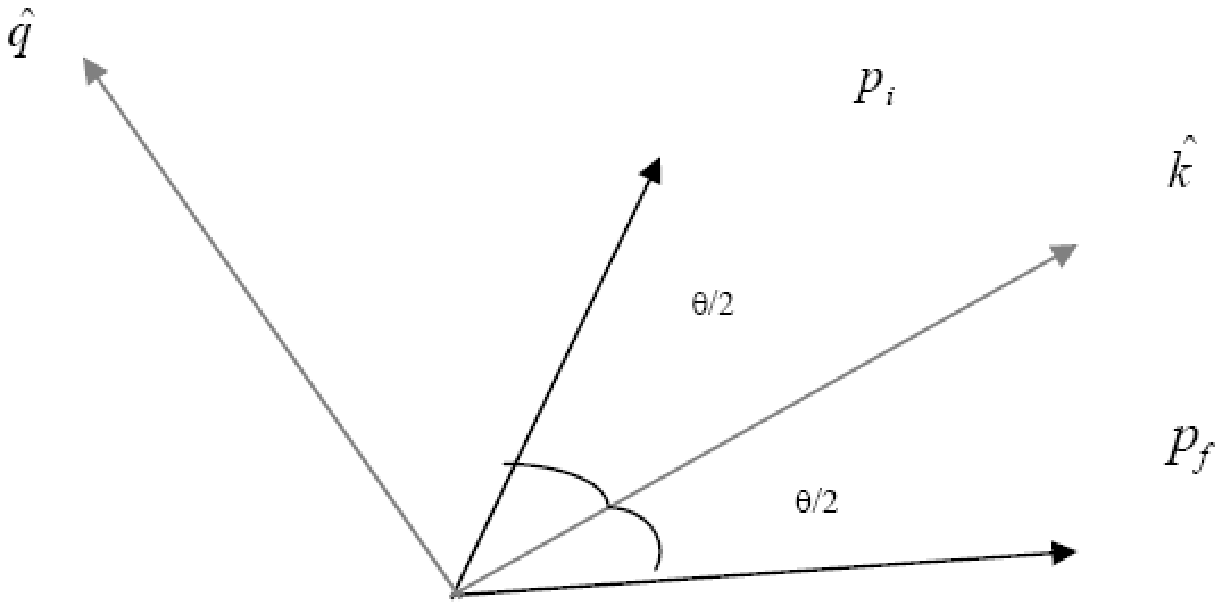}
\caption{\label{fig1} Scattering diagram in the xy-plane.}
\end{figure}
 Introducing a third unit vector
$\mathbf{\hat{l}}=\mathbf{\hat{k}}\times\mathbf{\hat{q}}$ that is
normal to the $\mathbf{\hat{k}}-\mathbf{\hat{q}}$ plane, we get a
set of three mutually orthogonal unit vectors which we employ to
define a new set of axes, see Figure 2 .  To this end, we
introduce the three operators
$\Sigma_k=\mathbf{\Sigma}.\mathbf{\hat{k}};\quad\Sigma_q=\mathbf{\Sigma}.\mathbf{\hat{q}}
$ and $\;\Sigma_l=\mathbf{\Sigma}.\mathbf{\hat{l}}$. Using the
identity $\mathbf{\Sigma}.\mathbf{A}\mathbf{\Sigma}.\mathbf{B}=
\mathbf{A}.\mathbf{B}+i\mathbf{\Sigma}.\mathbf{A}\times\mathbf{B}$
 we can immediately verify the following commutation and
 anti-commutation relations:
\begin{eqnarray}\label{11}
  \left[\Sigma_k,\Sigma_q\right] &=& 2i\Sigma_l \nonumber\\
\left[\Sigma_l,\Sigma_k\right] &=& 2i\Sigma_q\ \\
  \left[\Sigma_q,\Sigma_l\right] &=&
  2i\Sigma_k,\nonumber.
\end{eqnarray}

\begin{equation}\label{12}
\left\{\Sigma_l,\Sigma_k\right\} =
\left\{\Sigma_q,\Sigma_l\right\}=
\left\{\Sigma_q,\Sigma_k\right\}=0
\end{equation}
Thus, the consequences:
\begin{equation}\label{13}
(\Sigma_k)^2=(\Sigma_q)^2=(\Sigma_l)^2=I.
\end{equation}
and ,
\begin{equation}\label{13a}
i\Sigma_k=\Sigma_q\Sigma_l,\quad i\Sigma_q=\Sigma_l\Sigma_k,\quad
i\Sigma_l=\Sigma_k\Sigma_q
\end{equation}
\begin{figure}[htbp]
\includegraphics[width=8cm]{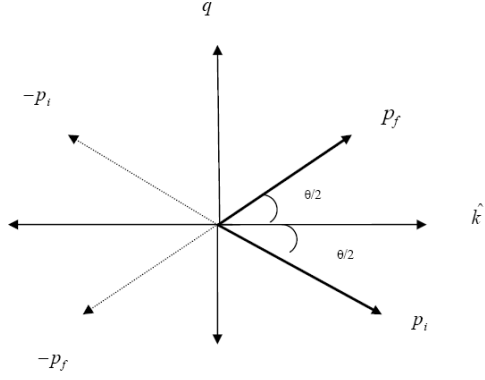}
\caption{\label{fig2} Scattering diagram in the $k-q$ plane.}
\end{figure}
 The above relations says that the newly introduced $\Sigma$
matrices furnish a representation of the $SU(2)$ algebra,and are
generators of rotation in the spin space.We will now express all
the spin operators and the SI in terms of these generators. We
will thus, demonstrate that the description of the
helicity-conserving first order transition in the spin space
becomes more symmetric. To start with, express
$\mathbf{\Sigma}.\mathbf{\hat{p}}_i$ and
$\mathbf{\Sigma}.\mathbf{\hat{p}}_f$ in terms of $\Sigma_k$ and
$\Sigma_q$ ( see Figure 2 ):
\begin{eqnarray}\label{13b}
  \label{15}
\mathbf{\Sigma.\hat{p_i}}&=&\cos\frac{\theta}{2}\Sigma_k-\sin\frac{\theta}{2}\Sigma_q\nonumber\\
\mathbf{\Sigma.\hat{p_f}}&=&\cos\frac{\theta}{2}\Sigma_k+\sin\frac{\theta}{2}\Sigma_q
\end{eqnarray}
The symmetry in the above expression between the helicity
operators of the initial and final particles - which goes with the
symmetry in Figure  - is obvious. One can actually go further and
check that - as the figure also suggests-
$\mathbf{\Sigma.\hat{p_i}}$ and $\mathbf{\Sigma.\hat{p_f}}$ are
related by a rotation about the $\mathbf{l}$-axis:
\begin{equation}\label{15b}
 \mathbf{\Sigma.\hat{p_f}}=U^{-1}(l,\theta)\mathbf{\Sigma.\hat{p_i}}U(l,\theta)
\end{equation}
The above equation makes explicit the intuitive picture that the
spin of the incident particle gets rotated by an angle $\theta $
to remain aligned along the direction of the momentum.
\section{The Transition in the $ \hat k-\hat q $ Basis}
In this section we will express the SI in terms of the newly
introduced generators and investigate the interesting consequences
of this. We will then write the scattering states in terms of the
$\mathbf{\hat k}$-basis and obtain an expression for the matrix
element in terms of these basis. We first note the following major
relations which can be easily  proven using
Eqs.(\ref{11})-(\ref{15})  :
\begin{eqnarray}
  \mathbf{\Sigma.\hat{p_f}}\Sigma_k \mathbf{\Sigma.\hat{p_i}} &=& \Sigma_k \label{15d}\\
  \mathbf{\Sigma.\hat{p_f}}\Sigma_q \mathbf{\Sigma.\hat{p_i}} &=&-\Sigma_q\label{15e}
\end{eqnarray}
Note how the above two equations go with the symmetry in Figure 2.
Now, from Figure 3, we have the unit vector $\mathbf{\hat a}$
appearing in the SI given as :
\begin{figure}[htbp]
\includegraphics[width=8cm]{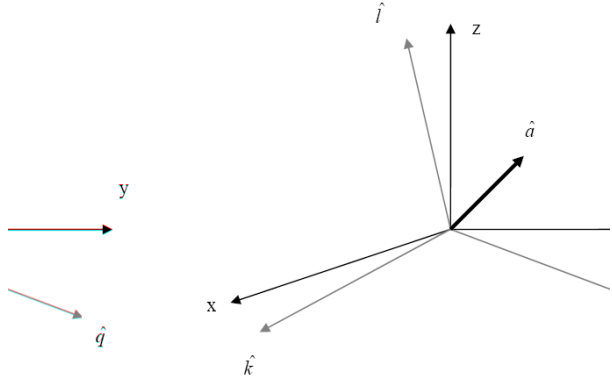}
\caption{\label{fig3} The components of $\mathbf{\hat a}$ in the
$k-q$ plane.}
\end{figure}
 \begin{eqnarray}\label{15c}
   \mathbf{\hat a} &=&\left(\mathbf{\hat a . \hat l}\right)\mathbf{\hat l}+\left(\mathbf{\hat a . \hat k}\right)\mathbf{\hat
   k}
   +\left(\mathbf{\hat a . \hat q}\right)\mathbf{\hat q} \\
    &=& A\mathbf{\hat l}+B\mathbf{\hat k}+C\mathbf{\hat q}
 \end{eqnarray}
 The spin interaction operator will then take the form:
\begin{equation}\label{16}
\gamma_5\mathbf{\Sigma}.\mathbf{\hat{a}}=A\gamma_5\Sigma_l
+B\gamma_5\Sigma_k+C\gamma_5\Sigma_q
\end{equation}
with $A,\:B$ and $C$ defined in Eq.(\ref {15c}) above. The
transition matrix element, Eq.(\ref{5c}), upon employing the
expansion given by Eq.(\ref{16})above can be further reduced. To
do this, consider first the matrix element of $\Sigma_q$, namely
$<\mathbf{\hat{p_f}};\pm|\gamma_5\Sigma_q|\mathbf{\hat{p_i}};\pm>$.
This can be written ( just by noting  that the states are
eigenstates of the initial and final helicity operators) as :
\begin{eqnarray}\label{17}
<\mathbf{\hat{p_f}};\pm|\gamma_5\Sigma_q|\mathbf{\hat{p_i}};\pm>&=&
<\mathbf{\hat{p_f}};\pm|\gamma_5\mathbf{\Sigma.\hat{p_f}}\Sigma_q
\mathbf{\Sigma.\hat{p_i}}|\mathbf{\hat{p_i}};\pm>\nonumber\\
&=&-<\mathbf{\hat{p_f}};\pm|\gamma_5\mathbf{\Sigma.\hat{p_f}}\Sigma_q
\mathbf{\Sigma.\hat{p_i}}|\mathbf{\hat{p_i}};\pm>\nonumber\\
\end{eqnarray}
Where we have used Eq.(\ref{15e})to write the second line. Letting
operators  act on their eigenstates  and noting that $\gamma_5$
commutes with all the $\Sigma 's$  , we get
\begin{equation}\label{18}
<\mathbf{\hat{p_f}};\pm|\gamma_5\Sigma_q|\mathbf{\hat{p_i}};\pm>=
-<\mathbf{\hat{p_f}};\pm|\gamma_5\Sigma_q|\mathbf{\hat{p_i}};\pm>
\end{equation}
with the obvious consequence:
\begin{equation}\label{19}
<\mathbf{\hat{p_f}};\pm|\gamma_5\Sigma_q|\mathbf{\hat{p_i}};\pm>=0
\end{equation}
The $\Sigma_q$ part of the SI does not contribute to the
helicity-conserving transition. This should not be too surprising,
as it is a guarantee of the gauge-invariance of the transition
probability. Obviously, under a gauge transformation
$\mathbf{A(q)}\longrightarrow \mathbf{A(q)}+\mathbf{q}
f(\mathbf{q})$, with  $f(\mathbf{q})$ arbitrary. So , if the
matrix element is to be gauge-invariant, which is indeed so, then
the contribution of $\Sigma_q$ should vanish.  We now move to the
$\Sigma_l$ matrix element. This, again, can be expressed as :
\begin{equation}\label{20}
<\mathbf{\hat{p_f}};\pm|\gamma_5\Sigma_l|\mathbf{\hat{p_i}};\pm>=
<\mathbf{\hat{p_f}};\pm|\gamma_5\mathbf{\Sigma.\hat{p_f}}\Sigma_l
\mathbf{\Sigma.\hat{p_i}}|\mathbf{\hat{p_i}};\pm>
\end{equation}
This can be reduced (see the appendix) to :
\begin{equation}\label{21}
<\mathbf{\hat{p_f}};\pm|\gamma_5\mathbf{\Sigma.\hat{p_i}}\Sigma_l
\mathbf{\Sigma.\hat{p_i}}|\mathbf{\hat{p_i}};\pm>=\pm
i<\mathbf{\hat{p_f}};\pm|\gamma_5(-\cos\frac{\theta}{2} \Sigma_q
+\sin\frac{\theta}{2} \Sigma_k)|\mathbf{\hat{p_i}};\pm>
\end{equation}
 The matrix element of the $\Sigma_q$ component vanishes as we have demonstrated above,
 and we are left with the $\Sigma_k$ contribution. Thus,putting every thing together we have
 the result:
 \begin{equation}\label{22}
<\mathbf{\hat{p_f}};\pm|\gamma_5\mathbf{\Sigma}.\mathbf{\hat{a}}|\mathbf{\hat{p_i}};\pm>=
(B\pm
iA\sin\frac{\theta}{2})<\mathbf{\hat{p_f}};\pm|\gamma_5\Sigma_k|\mathbf{\hat{p_i}};\pm>
\end{equation}
The transition is induced solely by $\gamma_5 \Sigma_k$, i.e the
component of the spin interaction  operator along the direction of
the total momentum vector $\mathbf{k}$ ! To see what is special
with this direction, look again at Figure 2. The
helicity-conserving transition is a transition that leaves the
component of the spin along $\mathbf{\hat k}$ invariant, while
flipping the component along  $\mathbf{\hat q}$. This is what
Eqs.(\ref{15d}) and (\ref{15e}) also say. Therefore, formulated in
the $ \hat k-\hat q $ basis, the conservation of helicity at first
order scattering in a static magnetic field amounts to the
conservation of the spin component along $\mathbf{\hat k}$ in the
transition and the flipping of the component along $\mathbf{\hat
q}$ . This is just what happens to the momentum of a classical
object; a ball say, as it bounces off a wall. The momentum along
the wall is conserved, while that parallel to it flips. In our
case, the "wall" is defined by the total momentum vector
$\mathbf{k}$, see Figure 4. The transition, however, takes place
in the spin space, and the relevant quantity is the orientation of
the spin of the particle !.
\begin{figure}[htbp]
\includegraphics[width=8cm]{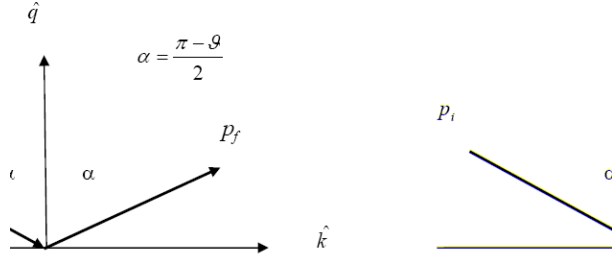}
\caption{\label{fig4} The bouncing ball picture of helicity
conservation}
\end{figure}
This picture can be enhanced by expanding the initial and final
helicity states in terms of the eigenstates of $\Sigma_k$ and
$\Sigma_q$, which can be achieved by simple rotations about the
$\mathbf{\hat l}-$axis. We focus here on states with positive
helicity; those with negative helicity can be obtained in exactly
the same manner. Indeed, from Figure 2, we can see that :
\begin{eqnarray}\label{24}
|\mathbf{\hat{p_i}};+>&=&U(\mathbf{ \hat
l},\frac{-\theta}{2})|\mathbf{\hat{k}};+>
=\cos\frac{\theta}{4}|\mathbf{\hat{k}};+>+ \sin
\frac{\theta}{4}|\mathbf{\hat{k}};->\nonumber\\
&=&U(\mathbf{ \hat l},-\frac{\theta+\pi }{2})|\mathbf{\hat{q}};+>
=\cos\frac{\theta+\pi }{4}|\mathbf{\hat{q}};+>+\sin \frac{\theta +
\pi}{4}|\mathbf{\hat{q}};->
\end{eqnarray}
and,
\begin{eqnarray}\label{25}
|\mathbf{\hat{p_f}};+>&=&U(\mathbf{ \hat
l},\frac{\theta}{2})|\mathbf{\hat{k}};+>
=\cos\frac{\theta}{4}|\mathbf{\hat{k}};+>- \sin
\frac{\theta}{4}|\mathbf{\hat{k}};->\nonumber\\
&=&U(\mathbf{ \hat l},\frac{\theta-\pi }{2})|\mathbf{\hat{q}};+>
=\sin\frac{\theta+\pi }{4}|\mathbf{\hat{q}};+>+\cos\frac{\theta +
\pi}{4}|\mathbf{\hat{q}};->
\end{eqnarray}
Investigating the above equations it is obvious that
\begin{eqnarray}
|<\mathbf{\hat{k}};\pm|\mathbf{\hat{p_i}};+>|^2=|<\mathbf{\hat{k}};\pm|\mathbf{\hat{p_f}};+>|^2
\end{eqnarray}
while,
\begin{eqnarray}
|<\mathbf{\hat{q}};\pm|\mathbf{\hat{p_i}};+>|^2=|<\mathbf{\hat{q}};\mp|\mathbf{\hat{p_f}};+>|^2
\end{eqnarray}

 In fact, one can check directly that the SI interaction connects initial and final $\mathbf{\hat{k}}-$ states
 with the same helicity only, i.e. no flip,but different helicity
 $\mathbf{\hat{q}}-$states.
 To see this ,we consider the matrix elements
$<\mathbf{\hat{k}};\mp|\gamma_5\Sigma_k|\mathbf{\hat{k}};\pm>$ and
$<\mathbf{\hat{q}};\pm|\gamma_5\Sigma_k|\mathbf{\hat{q}};\pm>$ and
show that they both vanish. Consider the first one :
\begin{eqnarray}\label{26}
<\mathbf{\hat{k}};\mp|\gamma_5\Sigma_k|\mathbf{\hat{k}};\pm>&=&
\pm<\mathbf{\hat{k}};\mp|\gamma_5|\mathbf{\hat{k}};\pm>\nonumber\\
&=&<\mathbf{\hat{k}};\mp|(\mp)\Sigma_k\gamma_5
(\pm)\Sigma_k|\mathbf{\hat{k}};\pm>\nonumber\\
&=&-<\mathbf{\hat{k}};\mp|\gamma_5|\mathbf{\hat{k}};\pm>
\end{eqnarray}
Thus,
\begin{equation}\label{27}
<\mathbf{\hat{k}};\mp|\gamma_5\Sigma_k|\mathbf{\hat{k}};\pm>=
\pm<\mathbf{\hat{k}};\mp|\gamma_5|\mathbf{\hat{k}};\pm>=
\mp<\mathbf{\hat{k}};\mp|\gamma_5|\mathbf{\hat{k}};\pm>=0
\end{equation}
Similarly,
\begin{eqnarray}\label{28}
<\mathbf{\hat{q}};\pm|\gamma_5\Sigma_k|\mathbf{\hat{q}};\pm>\nonumber
&=& <\mathbf{\hat{q}};\pm|\Sigma_q\gamma_5\Sigma_k
\Sigma_q|\mathbf{\hat{q}};\pm>\nonumber\\
&=& -<\mathbf{\hat{q}};\pm|\gamma_5 \Sigma_k
|\mathbf{\hat{q}};\pm>\nonumber
\end{eqnarray}
where in the last line we noted that $\Sigma_k$ and $\Sigma_q$
anticommute in view of Eqs.(\ref{12}).So, again:
\begin{equation}\label{29}
<\mathbf{\hat{q}};\pm|\gamma_5 \Sigma_k |\mathbf{\hat{q}};\pm>=0
\end{equation}
These results support our earlier arguments regarding the
conservation of the  $\mathbf{\hat k}$ component and the flipping
of the  $\mathbf{\hat q}$ component of the spin of the incident
particle.\\
Finally, one can, by expanding the initial and final states in
terms of the $\Sigma_k$ eigenstates, thus eliminating any
reference to these in the matrix element,  express the matrix
element totally in $\mathbf{\hat k}$ variables and states.
Starting from Eq.(\ref{22}), we express the matrix element (see
Eq.(\ref{24})and (\ref{25})) as:
\begin{equation}\label{30}
<\mathbf{\hat{p_f}};\pm|\gamma_5\Sigma_k|\mathbf{\hat{p_i}};\pm>=
<\mathbf{\hat{k}};\pm|U^{-1}(\mathbf{ \hat
l},\frac{\theta}{2})\gamma_5\Sigma_k U(\mathbf{ \hat
l},-\frac{\theta}{2})|\mathbf{\hat{k}};\pm>
\end{equation}
One can easily check that
\begin{equation}\label{31}
U^{-1}(\mathbf{ \hat l},\frac{\theta}{2})\gamma_5\Sigma_k
U(\mathbf{ \hat l},-\frac{\theta}{2})= \gamma_5\Sigma_k
\end{equation}
Combining this result with Eq.(\ref{22})we get
\begin{equation}\label{32}
<\mathbf{\hat{p_f}};\pm|\gamma_5\mathbf{\Sigma}.\mathbf{\hat{a}}|\mathbf{\hat{p_i}};\pm>=
(B\pm
iA\sin\frac{\theta}{2})<\mathbf{\hat{k}};\pm|\gamma_5\Sigma_k
|\mathbf{\hat{k}};\pm>
\end{equation}
Acting with $\Sigma_k$ on its eigenstates, we get the result:
\begin{equation}\label{32b}
<\mathbf{\hat{p_f}};\pm|\gamma_5\mathbf{\Sigma}.\mathbf{\hat{a}}|\mathbf{\hat{p_i}};\pm>=
\pm(B\pm iA\sin\frac{\theta}{2})<\mathbf{\hat{k}};\pm|\gamma_5
|\mathbf{\hat{k}};\pm>
\end{equation}
In the above equation, the only reference to the initial and final
states is through the kinematical/geometrical factors $A$ and $B$.
So, to calculate the transition matrix element for any vector
potential, just find these factors - which is a trivial task- and
plug them into the above expression. Things can be even further
simplified if we use the explicit forms of the spinors  :
\begin{equation}\label{33}
|\mathbf{\hat{k}};\pm>= N'\left(\begin{array}{c}
  \chi_\pm \\
  \frac{\sigma .\mathbf{k_0}}{E+m}\chi_\pm  \\
\end{array}\right)
\end{equation}
where $\chi_\pm $ are eigenstates of $\sigma.\mathbf{k_0}$ with
eigenvalues $\pm 1$, and $\mathbf{k_0}=p\mathbf{\hat k_0}$ is a
vector along $\mathbf{\hat k_0}$  with $p$ being the conserved
magnitude of the initial and the final momenta. Plugging this
expression into Eq.(\ref{32b}) and using
$\gamma_5=\left(\begin{array}{cc}
  0 & I \\
  I & 0 \\
\end{array}\right)$ ,
we have:
\begin{equation}\label{32c}
<\mathbf{\hat{p_f}};\pm|\gamma_5\mathbf{\Sigma}.\mathbf{\hat{a}}|\mathbf{\hat{p_i}};\pm>=
\pm(B\pm iA\sin\frac{\theta}{2})\left(\frac{2N'^2 p }{E+m}\right)
\end{equation}
This is just a "plug and play" formula, where one just fixes the
geometrical factors $A$ and $B$ for the specific vector potential
present , and then gets the spin sector of the matrix element
immediately. The following two examples illustrate this
explicitly.

\section{Examples}
In this section we consider two concrete examples of vector
potentials whose field configurations conserve helicity, and we
bring the  first order transition matrix elements of Dirac
particles in these potentials  to the form given by
Eq.(\ref{32b}). Consider first the Ahronov-Bohm (AB) potential
\cite{AB} which gives rise to a $\delta$-function mgnetic field
extended along the z-axis. This vector potential is given as:
\begin{equation}\label{33}
\mathbf{A(r)}=\frac{\Phi}{2\pi}
\frac{-y\hat{\mathbf{x}}+x\hat{\mathbf{y}}}{x^2+y^2}=\frac{\Phi}{2\pi\rho}\hat{\epsilon}_\varphi,
\end{equation}
where $\rho=\sqrt{x^2+y^2}$, $\hat{\epsilon}_\varphi$ is the unit
vector in the $\varphi$-direction, and $\Phi$ is the flux through
the AB tube. Since the magnetic field is along the z-axis; the
z-component of the incident momentum doe not change during the
scattering process . Therefore, we consider normal scattering,
i.e. take the incident, and consequently, the outgoing momenta to
be in the x-y plane. In such a geometry, $\mathbf{\hat l}$ is just
$\mathbf{\hat z}$. Pluggingb this vector potential into
Eq.(\ref{5c}), we get :
\begin{equation}\label{34}
S_{fi}^{(1)}=- 2\pi e|N|^{2} \delta (E_f-E_i) u^{\dagger} _f
\left( p_f,s_f \right)\left((-e\Phi) \frac{\alpha_1 q_2-\alpha_2
q_1}{q^2} \right)u _i \left(p_i,s_i \right).
\end{equation}
So,
\begin{equation}\label{35}
\mathbf{A(q)}=-\Phi \frac{q_2\mathbf{\hat x}-q_1\mathbf{\hat
y}}{q^2}=\frac{-\Phi}{q}\mathbf{\hat a(q)}
\end{equation}
with $\mathbf{\hat a(q)}$ given as
\begin{equation}\label{36}
\mathbf{\hat a(q)}=\frac{q_2\mathbf{\hat x}-q_1\mathbf{\hat y}}{q}
\end{equation}
For the purpose of applying the formula (\ref{32b}), we need to
find the geometrical factors $A$ and $B$. Obviously, $A=0$. As for
B, we note that we can  without any loss of generality, take the
incident momentum to be along the x-axis;  $\mathbf{
p_i}=p\mathbf{\hat x}$ so that $\mathbf{p_f}=p
\left(\cos\frac{\theta}{2}\mathbf{\hat
x}+\sin\frac{\theta}{2}\mathbf{\hat y}\right)$. Straight forward
algebra shows that with such a choice of the incident momentum, we
get $\mathbf{\hat a(q)}=\mathbf{\hat k}$ , so that
$\gamma_5\mathbf{\Sigma}.\mathbf{\hat{a}}=\gamma_5\mathbf{\Sigma}.\mathbf{\hat{k}}$,
 meaning that $B=1$. The matrix element  for the AB
 potential then becomes \cite{helicity}:
 \begin{equation}
 \label{36}
 \mathcal{M}
=<\mathbf{\hat{p_f}};\pm|\gamma_5\mathbf{\Sigma}.\mathbf{\hat{a}}|\mathbf{\hat{p_i}};\pm>
=<\mathbf{\hat{p_f}};\pm|\gamma_5\mathbf{\Sigma}.\mathbf{\hat{k}}|\mathbf{\hat{p_i}};\pm>
=\pm <\mathbf{\hat{k}};\pm|\gamma_5|\mathbf{\hat{k}};\pm>
\end{equation}
We can even move to calculate the scattering cross section. The
unpolarized scattering cross section of a Dirac particle in the AB
field is given as \cite{vera,Shikakhwa1,boz2}:
\begin{equation}\label{37}
\frac{d\sigma}{d\theta}=\frac{e^2\Phi^2}{2\pi
p^3\\sin^2\frac{\theta}{2}}\frac{1}{2}\sum_{s_i,s_f=\pm}
|<\mathbf{\hat{p_f}};s_f|\gamma_5\mathbf{\Sigma}.\mathbf{\hat{a}}|\mathbf{\hat{p_i}};s_i>|^2
\end{equation}
where the summation is over the initial and final particles'
helicities. As a consequence of Eqs.(\ref{36}) we have
$<\mathbf{\hat{p_f}};-|\gamma_5\mathbf{\Sigma}.\mathbf{\hat{a}}|\mathbf{\hat{p_i}};->
=- <\mathbf{\hat{k}};+|\gamma_5|\mathbf{\hat{k}};+>
=-<\mathbf{\hat{p_f}};+|\gamma_5\mathbf{\Sigma}.\mathbf{\hat{a}}|\mathbf{\hat{p_i}};+>$.
So, using  Eq.(\ref{32}), and taking the normalization constant
$N'=\sqrt{\frac{E+m}{4m}}$ we get
\begin{equation}\label{38}
\frac{d\sigma}{d\theta}=\frac{e^2\Phi^2}{8\pi
p\\sin^2\frac{\theta}{2}}
\end{equation}
which is the well-known AB scatttering cross section of a Dirac
particle at first order \cite{vera}.

 The second example is the vector potential of a magnetic
dipole, and is less symmetric  as the resulting field is not,
contrary to the AB one, axial. The vector potential of the dipole
is given by \cite{Jackson}
\begin{equation}\label{}
\mathbf{A(r)}=\frac{\mathbf{\mu}\times\mathbf{r}}{r^3}
\end{equation}
where $\mathbf{\mu}$ is the magnetic moment. The Fourier transform
of the above vector potential is (up to a numerical factor)
 $\mathbf{A(q)}=\frac{\mathbf{\mu}\times\mathbf{q}}{q^2}$. Thus,
 the first order matrix element reads
 \begin{equation}\label{5}
 S_{fi}^{(1)}=- 2\pi e|N|^{2}|(\frac{1}{q^2})
\delta (E_f-E_i) u^{\dagger} _f \left( p_f,s_f \right)\left(
\gamma_5\mathbf{\Sigma}.
\frac{\mathbf{\mu}\times\mathbf{q}}{|\mathbf{\mu}\times\mathbf{q}|}\right)u
_i \left(p_i,s_i \right).
\end{equation}
Therefore $\mathbf{\hat
a}=\frac{\mathbf{\mu}\times\mathbf{q}}{|\mathbf{\mu}\times\mathbf{q}|}$
. The kinematical factors of Eq.(\ref{32}) are just
$A=\mathbf{\hat
l}.\frac{\mathbf{\mu}\times\mathbf{q}}{|\mathbf{\mu}\times\mathbf{q}|}$
and $B=\mathbf{\hat
k}.\frac{\mathbf{\mu}\times\mathbf{q}}{|\mathbf{\mu}\times\mathbf{q}|}$.
 which are straight forward to calculate; just specify $\mathbf{\mu}$ and $\mathbf{p_i}$.
 Therefore, the transition matrix element reads now:
 \begin{equation}\label{39}
 S_{fi}^{(1)}=\mp 2\pi e|N|^{2}|(\frac{1}{q^2})
\delta (E_f-E_i) (\mathbf{\hat
k}.\frac{\mathbf{\mu}\times\mathbf{q}}{|\mathbf{\mu}\times\mathbf{q}|}\pm
i\mathbf{\hat
l}.\frac{\mathbf{\mu}\times\mathbf{q}}{|\mathbf{\mu}\times\mathbf{q}|}
\sin\frac{\theta}{2})<\mathbf{\hat{k}};\pm|\gamma_5
|\mathbf{\hat{k}};\pm>
\end{equation}
The cross section can be calculated straight forwardly from the
above amplitude.

\section{Conclusions}
The spin interaction in the first order $S-$matrix of a Dirac
particle in a static magnetic field was investigated. Noting that
the total momentum vector $\mathbf{k}=\mathbf{p_f}+\mathbf{p_i}$
and the momentum transfer vector $\mathbf{q}=\mathbf{p_f-p_i}$ are
always perpendicular, we  suggested that the three unit vectors;
$\mathbf{\hat k},\; \mathbf{\hat q}$ and $\mathbf{\hat
l}\equiv\mathbf{\hat {k}\times\hat{ q}}$  defined an "intrinsic"
coordinate system, where the transition, and particularly,the
conservation of helicity, could be described in an alternative,
more symmetric formalism. The three generators
$\Sigma_k\equiv\mathbf{\Sigma.\hat{k}},\;\Sigma_q\equiv\mathbf{\Sigma.\hat{q}}$,
and $\Sigma_l\equiv\mathbf{\Sigma.\hat{l}}$ were shown to close
the $SU(2)$ algebra. When the spin interaction operator
$\gamma_5\mathbf{\Sigma.\hat{a}}$ was  written in terms of these
generators, we have been able to reduce the transition in the spin
space to an expression proportional to the matrix element of the
operator $\gamma_5\Sigma_k$. \\
Expressing $\mathbf{\Sigma.\hat{p_i}}$ and
$\mathbf{\Sigma.\hat{p_f}}$ and their eigenstates in terms of
$\Sigma_k,\;\Sigma_q$, and their eigenstates, we have demonstrated
that the conservation of helicity can be formulated as the
invariance of the $\mathbf{\hat k}$ component of the spin of the
particle and the flipping of its $\mathbf{\hat q}$ component. An
intuitive physical picture of the transition, similar to that of a
ball bouncing off a wall was suggested. The scattering matrix
element was  written, for any static field configuration, as the
matrix element of the $\gamma_5\Sigma_k$ in $\Sigma_k$ basis,
multiplied by kinematical/geometrical factors which carry the only
reference to the initial and final momenta.

\begin{appendix}\textbf{Appendix}\\
We show here how to derive Eqs.(\ref{21}) in the text. We start
with
\begin{equation}\label{42}
<\mathbf{\hat{p_f}};\pm|\gamma_5\Sigma_l|\mathbf{\hat{p_i}};\pm>=
<\mathbf{\hat{p_f}};\pm|\gamma_5\mathbf{\Sigma.\hat{p_f}}\Sigma_l
\mathbf{\Sigma.\hat{p_i}}|\mathbf{\hat{p_i}};\pm>=
\end{equation}
Look at :
\begin{equation}\label{40}
\mathbf{\Sigma.\hat{p_f}}\Sigma_l
\mathbf{\Sigma.\hat{p_i}}|\mathbf{\hat{p_i}};\pm>=
\mathbf{\Sigma.\hat{p_f}}\Sigma_l(\cos\frac{\theta}{2} \Sigma_k
-\sin\frac{\theta}{2} \Sigma_q)|\mathbf{\hat{p_i}};\pm>
\end{equation}
Using Eqs.(\ref{13a}), this can be written as:
\begin{equation}\label{41}
\mathbf{\Sigma.\hat{p_f}}\Sigma_l
\mathbf{\Sigma.\hat{p_i}}|\mathbf{\hat{p_i}};\pm>=
i\mathbf{\Sigma.\hat{p_f}}(\cos\frac{\theta}{2} \Sigma_q
+\sin\frac{\theta}{2} \Sigma_k)|\mathbf{\hat{p_i}};\pm>
\end{equation}
Eqs.(\ref{15d}) and (\ref{15e}) allow us to re-introduce
$\mathbf{\Sigma.\hat{p_i}}$ and thus bring this into the form
\begin{equation}\label{42}
\mathbf{\Sigma.\hat{p_f}}\Sigma_l
\mathbf{\Sigma.\hat{p_i}}|\mathbf{\hat{p_i}};\pm>=
i(-\cos\frac{\theta}{2} \Sigma_q
\mathbf{\Sigma.\hat{p_i}}+\sin\frac{\theta}{2}
\Sigma_k\mathbf{\Sigma.\hat{p_i}})|\mathbf{\hat{p_i}};\pm>
\end{equation}
Allowing the operator $\mathbf{\Sigma.\hat{p_i}}$ to act on its
eigenstates, we get:
\begin{equation}\label{42}
\mathbf{\Sigma.\hat{p_f}}\Sigma_l
\mathbf{\Sigma.\hat{p_i}}|\mathbf{\hat{p_i}};\pm>= \pm
i(-\cos\frac{\theta}{2} \Sigma_q +\sin\frac{\theta}{2}
\Sigma_k)|\mathbf{\hat{p_i}};\pm>
\end{equation}
Our result, now, follows immediately;
\begin{equation}\label{43}
<\mathbf{\hat{p_f}};\pm|\gamma_5\Sigma_l
|\mathbf{\hat{p_i}};\pm>=\pm
i<\mathbf{\hat{p_f}};\pm|\gamma_5(-\cos\frac{\theta}{2} \Sigma_q
+\sin\frac{\theta}{2} \Sigma_k)|\mathbf{\hat{p_i}};\pm>
\end{equation}

\end{appendix}
\bibliography{helicity}

\begin{thebibliography}{8}
\expandafter\ifx\csname natexlab\endcsname\relax\def\natexlab#1{#1}\fi
\expandafter\ifx\csname bibnamefont\endcsname\relax
  \def\bibnamefont#1{#1}\fi
\expandafter\ifx\csname bibfnamefont\endcsname\relax
  \def\bibfnamefont#1{#1}\fi
\expandafter\ifx\csname citenamefont\endcsname\relax
  \def\citenamefont#1{#1}\fi
\expandafter\ifx\csname url\endcsname\relax
  \def\url#1{\texttt{#1}}\fi
\expandafter\ifx\csname urlprefix\endcsname\relax\def\urlprefix{URL }\fi
\providecommand{\bibinfo}[2]{#2}
\providecommand{\eprint}[2][]{\url{#2}}

\bibitem[{\citenamefont{M.Jacob and G.C.Wick}(1959)}]{Jacob}
\bibinfo{author}{\bibnamefont{M.Jacob}} \bibnamefont{and}
  \bibinfo{author}{\bibnamefont{G.C.Wick}}, \bibinfo{journal}{Ann.Phys.}
  \textbf{\bibinfo{volume}{7}}, \bibinfo{pages}{404} (\bibinfo{year}{1959}).

\bibitem[{\citenamefont{J.J.Sakurai}(1967)}]{sakurai}
\bibinfo{author}{\bibnamefont{J.J.Sakurai}}, \emph{\bibinfo{title}{Advanced
  Quantum Mechanics}} (\bibinfo{publisher}{Addison-Wesley, Massachusettes},
  \bibinfo{year}{1967}).

\bibitem[{\citenamefont{Y.Aharonov and D.Bohm}(1959)}]{AB}
\bibinfo{author}{\bibnamefont{Y.Aharonov}} \bibnamefont{and}
  \bibinfo{author}{\bibnamefont{D.Bohm}}, \bibinfo{journal}{Phys. Rev.}
  \textbf{\bibinfo{volume}{115}}, \bibinfo{pages}{485} (\bibinfo{year}{1959}).

\bibitem[{\citenamefont{A.Albeed and M.S.Shikakhwa}(2008)}]{helicity}
\bibinfo{author}{\bibnamefont{A.Albeed}} \bibnamefont{and}
  \bibinfo{author}{\bibnamefont{M.S.Shikakhwa}},
  \bibinfo{journal}{Int.J.Theor.Phys.} \textbf{\bibinfo{volume}{47}},
  \bibinfo{pages}{2748} (\bibinfo{year}{2008}).

\bibitem[{\citenamefont{F.Vera and I.Schmidt}(1990)}]{vera}
\bibinfo{author}{\bibnamefont{F.Vera}} \bibnamefont{and}
  \bibinfo{author}{\bibnamefont{I.Schmidt}}, \bibinfo{journal}{Phys.Rev. D}
  \textbf{\bibinfo{volume}{42}}, \bibinfo{pages}{3591} (\bibinfo{year}{1990}).

\bibitem[{\citenamefont{M.S.Shikakhwa and N.K.Pak}(2003)}]{Shikakhwa1}
\bibinfo{author}{\bibnamefont{M.S.Shikakhwa}} \bibnamefont{and}
  \bibinfo{author}{\bibnamefont{N.K.Pak}}, \bibinfo{journal}{Phys.Rev.D}
  \textbf{\bibinfo{volume}{67}}, \bibinfo{pages}{105019}
  (\bibinfo{year}{2003}).

\bibitem[{\citenamefont{M.Boz and N.K.Pak}(2000)}]{boz2}
\bibinfo{author}{\bibnamefont{M.Boz}} \bibnamefont{and}
  \bibinfo{author}{\bibnamefont{N.K.Pak}}, \bibinfo{journal}{Phys.Rev.D}
  \textbf{\bibinfo{volume}{62}}, \bibinfo{pages}{045022}
  (\bibinfo{year}{2000}).

\bibitem[{\citenamefont{J.D.Jackson}(1975)}]{Jackson}
\bibinfo{author}{\bibnamefont{J.D.Jackson}}, \emph{\bibinfo{title}{Classical
  Electrodynamics}} (\bibinfo{publisher}{second edition,John Wiley and Sons},
  \bibinfo{year}{1975}).

\end{thebibliography}

\end{document}